\documentstyle[12pt,aasms4]{article}

\def\deg{$^{\rm o}$}

\received{}
\accepted{}

\lefthead{Crenshaw et al.}
\righthead{Kinematics in NGC 4151}

\begin{document}

\title{Kinematics of the Narrow-Line Region in the Seyfert 2 Galaxy Mrk 3\altaffilmark{1}}
\author{Jose R. Ruiz\altaffilmark{2},
D.M. Crenshaw\altaffilmark{2,3},
S.B. Kraemer\altaffilmark{2,3},
G.A. Bower\altaffilmark{4},
T.R. Gull\altaffilmark{3},
J.B. Hutchings\altaffilmark{5},
M.E. Kaiser\altaffilmark{6},
\& D. Weistrop\altaffilmark{7}}

\altaffiltext{1}{Based on observations made with the NASA/ESA Hubble Space 
Telescope. STScI is operated by the Association of Universities for Research in 
Astronomy, Inc. under NASA contract NAS5-26555.}
\altaffiltext{2}{Catholic University of America}
\altaffiltext{3}{Laboratory for Astronomy and Solar Physics, NASA's Goddard
Space Flight Center, Code 681, Greenbelt, MD 20771}
\altaffiltext{4}{NOAO, 950 N. Cherry Street, Tucson, AZ 85726-6732}
\altaffiltext{5}{Dominion Astrophysical Observatory, National Research Council of
Canada, 5071 West Saanich Rd., Victoria, BC V8X 4M6, Canada}
\altaffiltext{6}{Department of Physics \& Astronomy, Johns Hopkins University,
3400 North Charles St., Baltimore, MD 21218}
\altaffiltext{7}{Department of Physics, University of Nevada at Las Vegas, 
4505 South Maryland Parkway, Las Vegas, NV 89154-4002}

\begin{abstract}
We present measurements of radial velocities for the narrow-line region (NLR) 
gas in the Seyfert 2 galaxy Mrk 3 out to $\sim$1 kpc from the nucleus.  The
observations consist of two datasets, both using the Space Telescope Imaging 
Spectrograph (STIS) on board the $Hubble$ $Space$ $Telescope$ $(HST)$:
1) An [O~III] slitless spectrum with the G430M grating of the inner 3$''$ 
around the nucleus, and 2) a long-slit observation centered on the nucleus 
(PA = 71\deg) using the G430L grating and the 52$''$ x 0$''$.1 aperture.  
Our results produce radial velocity maps of the emission-line gas.  These maps
indicate general trends in the gas motion, which include: blueshifts and 
redshifts on either side of the nucleus, steep velocity rises from systemic up
to $\sim$$\pm$700 km s$^{-1}$ taking place in the inner 0$''$.3 (0.8 kpc) both
east and west of the nucleus, gradual velocity descents back to near-systemic
values from 0$''$.3-1$''$.0, slightly uneven velocity amplitudes on each
side of the nucleus, and narrow velocity ranges over the entire observed region.

When fitted to kinematic modeling programs for the NLR gas, the data clearly favor
a model where the gas exists in a partially filled bicone, is accelerated 
radially away from the nucleus, and is followed by a constant deceleration
(possibly due to collision with an ambient medium).  This geometry and general 
kinematic model is in agreement with previous work done on the NLR gas of NGC
1068 and NGC 4151.  On scales of hundreds of parsecs, we conclude that radial 
outflow may be a common feature of Seyfert galaxies.
\end{abstract}


\section{Introduction}
In many Seyfert 2s, the clouds appear to lie in a biconical or roughly linear 
configuration surrounding the nucleus (Schmitt \& Kinney 1996, etc).  Various 
kinematic models have been proposed to explain NLR cloud motion.  Capetti et al. 
(1995) have compared optical and radio measurements of the NLR of Mrk 3 and 
concluded that the NLR clouds are the result of radio jet plasma expanding away 
from the bicone axis. Winge et al. (1997), (1999) postulate gravitational motions 
for the NLR in NGC 4151.  Recently, Crenshaw \& Kraemer (2000) and Kaiser et al. 
(2000) have determined radial velocities as a function of position in the NLRs 
of NGC 1068 and NGC 4151 (the brightest Seyfert 2 and Seyfert 1, respectively) 
with the STIS on $HST$.  Crenshaw et al. (2000) have proposed a model where clouds 
on the surface of a bicone are radially accelerated from the nucleus by wind
pressure or radiation pressure, encounter and collide with an ambient medium, 
then decelerate to near-systemic values.  It explains the general trends seen 
in the radial velocity as a function of position in the inner kiloparsec around 
the nuclei of these galaxies.

Mrk 3 is a well-studied Seyfert 2 galaxy, which shows evidence for a hidden
Seyfert 1 nucleus from broad polarized emission lines (Schmidt \& Miller 1985). 
The host galaxy is classified as an elliptical or S0 galaxy type.  It lies 53 
Mpc away (H$_{0}$=75 km s$^{-1}$ Mpc$^{-1}$, z$=$0.0135, 3.82$''$ kpc$^{-1}$). It 
has been studied in every wavelength regime, from the X-ray (Griffiths, 1998; 
Georgantopoulos et al, 1999) through the radio (Kukula et al. 1999).  Mrk 3 has 
bright [O III] emission-line clouds that lie in a biconical configuration (apex 
of the two cones coincident with the nucleus) along PA$=$80\deg, with a 
half-opening angle of 22.5\deg\ (Schmitt \& Kinney 1996).  At the end of the 
western cone, a large, diffuse knot appears, while on the end of the eastern cone, 
a bright knot appears out of the bicone, giving the entire structure an `S' shape 
(Kukula et al. 1993).  Schmitt \& Kinney measure the clouds as extending 280 pcs 
on either side of the nucleus.  Recent X-ray observations of Mrk 3 using 
$Chandra$ (Sako et al. 2000) have found soft X-ray extended emission which lies 
along this PA.  Radio jets have also been observed (Axon et al. 1998, 
Kukula et al. 1999) along the same PA; these appear to follow the biconical structure, 
although the half-opening angle is far less ($\sim$~8-10\deg).  The jets also have 
an `S' shape (though much less pronounced) in the same regions as the [O~III] emission.  
Though they lie close to the emission-line clouds, they are not exactly coincident; 
Axon et al. (1998) suggest that they lie along the convex edge of the S-shaped curvature.  

A fainter set of clouds, situated along P.A.$\sim$100\deg, are seen further 
out in the ENLR.  These clouds extend from 1.$''$0 on either side of the nucleus 
to about 3$''$.2.  This group is more diffuse, fainter in surface brightness
by an order of magnitude, and follows the [O~III] emission contour lines seen in 
the ground-based observations of Pogge \& De Robertis (1993).  

In this paper, we present two $HST$ spectra (one long-slit, one slitless) of the 
Mrk~3 NLR clouds.  They provide consistent values of radial velocities as functions 
of NLR cloud positions.  The two datasets are then fitted to a kinematic modeling 
program that provides a radial velocity map of clouds within a bicone, given a 
velocity law that directs their motion.  When fitted to the radial acceleration + 
constant deceleration velocity law, the trends seen in both datasets are matched well.  
In \S 2 we detail the observations, while the data analysis is presented in \S 3.  
The results from the two datasets are given in \S 4.  The discussion of the model 
is given in \S 5, with \S 6 providing the overall discussion.  \S 7 presents the 
conclusions.

\section{Observations}
In order to find the NLR knots in the slitless observations, and hence, determine
their velocities, a companion [O III] image, as well as a continuum image are 
required.  To that end, the archival images of Mrk 3 were obtained.  The first of 
these is a Wide Field Planetary Camera (WFPC) [O III] observation.  This image 
served to match the bright NLR clouds.  A Wide Field Planetary Camera 2 (WFPC 2) 
continuum image was also retrieved, as well as a Faint Object Camera (FOC) [O III] 
image.  The FOC image is shown in Figure 1.  It served to match the faint ENLR 
clouds.  A summary of all the observations is given in Table 1.

The new slitless observations take advantage of STIS's spatial resolution (0$''$.1) 
and the G430M spectral resolution ($\lambda$/$\Delta$$\lambda$$\approx$10,000).  
The observations were centered at 5093 \AA, with a bandwidth of 286 \AA.  In addition 
to [O III] $\lambda$5007, [O III] $\lambda$4959 and H$\beta$ were also observed.  
The spectral region around [O III] $\lambda$5007 is shown in Fig. 2.  The horizontal 
axis is the dispersion axis, while the vertical axis is the spatial axis, as indicated 
by the scale.  The bright NLR clouds are smeared out along the dispersion axis, 
indicating large velocity dispersions.  They are also shifted along this axis, from 
which their radial velocities can be calculated.  The faint ENLR clouds are also 
seen with apparently smaller dispersions. 

We also obtained long-slit observations of Mrk 3 using the G140L, G230L, G430L 
and G750L gratings and the 52$''$ x 0$''$.1 aperture.  The complete observations 
covered the wavelength range from 1150-10,000 \AA.  A full analysis of these 
data will be presented separately (Collins et al. 2001, in preparation).  
For this work, the single emission line of [O~III] $\lambda$5007 from these 
observations was used.  The slit had a PA of 71\deg, and was chosen to pass 
through the nucleus.  The position of the long-slit is shown in Fig. 1, overlaid 
on the FOC image for comparison with the [O III] clouds.  

\section{Data Analysis}
The data reduction was done with IDL software developed for the STIS Instrument 
Definition Team at Goddard Space Flight Center (Lindler et al. 1999).  The spectral 
images were cleaned of cosmic rays during the image processing.  Once the archival 
images were retrieved, they were also cleaned of cosmic rays.  Determining radial 
velocities of NLR clouds using slitless spectra has been done previously using 
NGC 4151 (Hutchings et al. 1998, Kaiser et al. 2000) and NGC 1068 (Crenshaw et al. 2000).  
These authors describe the STIS slitless data analysis; a brief synopsis follows here.

The data analysis consists of matching [O III] undispersed clouds (from an FOC,
WFPC or WFPC2 image) with their counterparts in a STIS dispersed image.  In order 
to make the one-to-one correspondence between clouds, the non-STIS images must be 
rotated, aligned, and corrected for anamorphic magnification with respect to the
STIS image (spatial scale of 0$''$.0507 pixel$^{-1}$).  Once the spatial scale
is set, then the wavelength scale (ultimately a velocity scale) must be set using 
the calibration spectra.  To set the velocity scale, the STIS dispersed image is
aligned with the direct [O III] images such that there is no shift for a cloud at
the systemic redshift (z=0.0135, Tifft \& Cooke 1988) of Mrk 3.  The redshifted 
wavelength is taken from 21 cm measurements of H I, assumed to be at rest with 
respect to the central galactic region.  We note that the WFPC [O III] observation 
and WFPC2 continuum image were taken at different times.  Despite this, we were able 
to identify the bright, distinct emission line knots in both images.

Once the STIS dispersed image and the non-STIS undispersed image are aligned, 
Gaussian fits are made to the NLR knots in both images on each row parallel 
to the dispersion axis.  This is to accurately determine the positions and 
FWHMs for each knot.  The difference in position (in pixels) between an 
undispersed and dispersed knot is converted to a difference in \AA, then to
one in velocity.  The difference in width (in quadrature) gives the velocity 
dispersion of the emission knot.  For some purposes, the individual line 
measurements are averaged over each cloud in order to obtain a single value of 
radial velocity and dispersion, and the standard deviations are used as 
uncertainties. 

For the long-slit data, Gaussian fits were made to the NLR knots on each row 
perpendicular to the dispersion (in 0$''$.05 intervals).  Once the positions and 
FWHMs were calculated, the radial velocities and dispersions for each row could 
be calculated using the same fitting routines as for the slitless spectrum.  The 
FWHMs were corrected for the instrumental broadening of 300 km s$^{-1}$.  For the 
purpose of comparing slitless and long slit data, we determined which clouds lay 
in the slit from the PA of the slit.  We then averaged the long-slit radial 
velocities for these clouds over each individual cloud, thus obtaining two values 
of radial velocity from the two different methods and plotted them against each 
other.  For both sets of radial velocities, the standard deviations calculated 
were used for the uncertainties. Figure 3 shows averaged slitless radial velocities 
plotted against the corresponding averaged long-slit radial velocities.  The plot 
shows very good agreement, to within $\sim$50 km s$^{-1}$, indicating that there 
are no systematic errors with the slitless technique.   

\section{Results}
The entire dataset is seen in Figure 4, which shows a plot of radial velocity 
versus distance.  Each point represents an average over each cloud.  The inner 
region clouds are seen to have a wide range of radial velocities, from $-$1000 
km s$^{-1}$ to +600 km s$^{-1}$.  There are roughly equal numbers of clouds that 
show redshifts and blueshifts.  As noted in the previous section, the long-slit 
cloud values are consistent with the slitless spectral values, both in terms of 
location and velocity.  The fainter, ENLR clouds stretch for about 1$''$.5 
further out in either direction than the inner region clouds.  The fainter 
clouds show a narrower range of radial velocity, from +400 km s$^{-1}$ down to 
$-$200 km s$^{-1}$, with the majority of the clouds (75\%) showing redshifts.

Figure 5 shows the velocity distribution of all the clouds.  The NLR clouds are 
difficult to distinguish due to their proximity to each other, so we present this 
figure in order to show the ENLR clouds.  The eastern half of these clouds is 
exclusively redshifted, while the majority of the western half is so with lower 
magnitudes.  Figure 6 is an expanded view of the central region of Figure 5 to 
show the inner region clouds' velocity distribution in greater detail.  The inner 
region clouds, in contrast to the extended region clouds, have equal numbers of 
blueshifted and redshifted clouds on either side of the nucleus.  The blueshifted 
clouds to the east tend to have greater velocities, while to the west the velocities 
are similar for redshifts and blueshifts.  On either side, maximum values of 
radial velocities are reached $\sim$0$''$.3 away from the center for both sets 
of data.

Figure 7 shows the unbinned NLR radial velocities (relative to systemic) plotted 
against distance from the optical continuum center.  The unbinned velocities are
used for the model in order to take advantage of $HST's$ spatial resolution.  
Long-slit points are shown, along with the slitless spectral points that 
lie within the slit.  This set of points is used because the modeling program we
use (to be discussed in the next section) simulates a long-slit, so we choose only
those points within the slit.  The velocity errors are only measurement errors 
in this case, and are comparable to the size of the symbols.  The two sets of data 
points are seen to be compatible.  They show properties that must be duplicated 
by any model fit.  Now we briefly discuss each of these properties. 

Both sets of data points show fast rises in velocities (from systemic values at
the nucleus to $\sim$$\pm$600 km s$^{-1}$) out to $\sim$0$''$.25 of the nucleus 
on each side.  The climbs in velocities are seen in both redshifts and blueshifts. 
All the rises are then followed by shallower velocity downturns, so that at 
$\sim$0$''$.7-1$''$.0, the velocities have returned to near systemic values.  
The amplitudes of the maximum velocities are not equal.  The amplitudes range
from $\sim$300 km s$^{-1}$ on the blueshifted west side to $\sim$800 km s$^{-1}$ 
on the blueshifted east side (ignoring a few high-velocity points).  This variation 
in amplitude implies that any fitted cone is tilted, and in fact, the angle of 
inclination can be calculated by the amplitude difference.  Finally, the range of 
velocities is fairly narrow.  For example, $\sim$0$''$.30 west of the nucleus, 
blueshifted velocities are seen exclusively from -200 to -300 km s$^{-1}$, while 
redshifted velocities are seen spanning a narrow range from 300 to 500 km s$^{-1}$.  
At this distance, there are no velocities seen from -200 to 300 km s$^{-1}$. 
Any model fit must be able to match these narrow velocity ranges.  

\section{Model Fitting}
Once the spatial orientations and radial velocities of the clouds were obtained,
we attempted to fit these observations using kinematic modeling programs.  These 
programs calculate the radial velocities and spectral lines of material on the
surface of a thin disk, or a bicone, either filled or hollow.  Each geometry can 
assume one of various velocity laws that control the material's movement.  We 
concentrate here on the inner NLR out to $\sim$1$''$.0 on either side of the 
nucleus.  Later, we will discuss briefly the ENLR clouds and their motion.

We can immediately rule out gravitational rotation models by calculating 
the mass required to impart radial velocities on the order of 
500-1000 km s$^{-1}$ at a distance of 100-200 pcs away from the nucleus.
This mass is of the order of 10$^{9-10}$ M$_\odot$.  Typical masses for 
black holes in Seyfert nuclei are 10$^{6-8}$ M$_\odot$ (Peterson \& Wandel 2000).  
Observationally, the observed morphology of the NLR does not suggest a disk 
geometry, while the redshifts and blueshifts on either side of the nucleus 
cannot be the result of simple Keplerian rotation.  Thus we are left with outflow 
models, or models where material flows tangentially outward from the radio axis. 

The bicone program has been used previously to model the NLR emission-line 
clouds of NGC 1068 and NGC 4151 (Crenshaw \& Kraemer 2000, Crenshaw et al. 2000).
The two cones (one on either side of the nucleus) are assumed to possess identical 
properties, including geometry, size and velocity law.  In addition, the cones are 
assumed to have a filling factor of 1 within the minimum and maximum half opening
angle, and not to absorb [O~III] photons.  We adjust certain parameters, shown in 
Table 2, to obtain the best fit.  The program creates a two-dimensional velocity map, 
which is sampled through a simulated slit.  We applied the models to unbinned 
velocities in order to obtain the best spatial resolution.

Several model input variables can be constrained from the observations.  The first 
of these was the extent of the NLR.  Based on the approximate placing of a bicone
on the Mrk 3 NLR by Schmitt \& Kinney (1996), we measured its maximum extent as 
$\sim$0.75$''$.  In addition to a minimum and maximum distance (in pixels) of the 
cones, the program requires a minimum and maximum half-opening angle.  The maximum 
angle was measured from the images, giving a value of 25\deg.  This agrees with 
Schmitt \& Kinney (1996), who measure a maximum half-opening angle of 22.5\deg.  
The minimum half-opening angle is not visible in the [O III] images, so it was 
varied to match the data.  The optimum value for the models was 15\deg.  This value 
places the emission-line material outside the observed radio jet cone (half-opening
angle $\sim$7-8\deg) (Capetti et al. 1995).  

The inclination angle was calculated based on the differences between the radial 
velocity maxima on the W and E sides of the cones.  The maximum blueshifts are 
higher by $\sim$300 km s$^{-1}$ than the maximum redshifts on the east side.  
The NLR inclination angle was then calculated as $\sim$5-10\deg, using simple 
trigonometry.  Finally, the value for the maximum deprojected velocity of the NLR 
gas was chosen so that it would match the observed NLR radial velocity peak 
($\sim$-800 km s$^{-1}$).  The best fit parameters of all the models are shown 
in Table 3.  The results of the models are summarized below.

1. The radial acceleration (RA) outflow model consists of NLR clouds being driven 
away from the nucleus, perhaps by winds or jets.  The acceleration is along 
the bicones' entire length.  The best fit was not able to match the high 
velocities near the center, given the observed parameters.  The only way to
marginally match these velocities was to widen the half-opening angle 
past $\sim$40\deg, but the sharp downturns cannot then be fit.  It is clear from
the images that the ionization cone's half-opening angle cannot be more than 
$\sim$30\deg.  If there is acceleration along the bicone, it cannot take place
along the entire length of the NLR.

2. The constant velocity (CV) model consists of clouds with a negligible drag 
force, having been accelerated out to some distance (small compared to the NLR),
then proceeding with constant velocity.  This model is able to match the high central 
velocities $\sim$0$''$.3 from the nucleus.  Further out, however, the modeled 
velocities remain at a constant value out to the ENLR, whereas the observed 
velocities drop to near systemic values by $\sim$1$''$.0 out from the center.

3. The constant tangential (CT) model consists of NLR clouds moving radially away 
from the central radio axis.  This would be seen if the radio plasma expanded within 
the emission-line bicone.  This fit resembles the CV model, except that it predicts
equal magnitude redshifts and blueshifts on either side of the nucleus.  This is 
certainly not the case, as seen in the two datasets.  Note that this model predicts 
velocity magnitudes substantially less (1/3 to 1/6) than the other models (see Table 3).  
These velocities appear to be too low.  There are a number of other inconsistencies with
this model, which are discussed in \S 6. 

4. The model that fit the most data points is the radial acceleration + constant 
deceleration (RA+CD).  The model can be visualized as material first accelerated by 
wind or radiation pressure from the nucleus, which then impacts an ambient medium 
and then decelerates at a constant rate.  This model implies that the emission-line 
clouds originate from a region closer to the nucleus and move outward from there.
Figure 8 shows the long-slit and slitless data points overlaid with the shading from 
this model.  Obviously, this model does not perfectly fit every point, but it fits 
the gross features of the observations well.  Many of the discrepant points can be 
ascribed to slightly different acceleration or deceleration laws in different quadrants.  
The discrepant high-velocity points suggest clouds that perhaps do not encounter the 
ambient, possibly patchy medium, or encounter it in a region of lower density and 
do not decelerate as much.  
  		                                             
\section{Discussion} 
The slitless spectral method of determining radial velocity gives consistent 
values with the long slit method, as shown in Figures 3 and 7.  This result 
gives confidence in future work using the slitless method, and has been shown 
before for NGC 4151 (Hutchings et al. 1998, Crenshaw et al. 2000).  The best fit 
RA+CD model shows, in addition to fitting all of the trends seen in the data, 
some discrepancies.  These can best be explained by slightly different
acceleration/deceleration laws in different directions. 

While the NLR clouds are fitted most closely by the RA+CD model, the ENLR 
clouds require a different model.  Those clouds (with velocities 
$\lesssim$ 350 km s$^{-1}$) appear to be influenced by the gravitational 
potential of the supermassive black hole (SMBH) and inner galaxy, rather than 
the outflowing material.  This hypothesis agrees with surface photometry on Mrk 3 
done by one of us (Bower).  Ellipsoids were fit to the 
surface brightness of Mrk 3, from 0.01$''$ out to 100$''$.  From these 
ellipsoid fits, a spherical dynamical model was used to predict a rough upper 
limit on the rotational radial velocity induced by the gravitational potential. 
For the range from 1$''$ to $\sim$5$''$, where the ENLR clouds reside, the 
projected radial velocities are predicted to be $\lesssim$200 km s$^{-1}$.  
This heuristic result agrees roughly with the observed ENLR cloud velocities, 
although we cannot explain the preponderance of redshifted clouds with this model. 
They may be due to a lack of ionized gas at the positions that would produce 
blueshifts. 

The orientation of the host galaxy has been previously reported as 27\deg\ out of
the plane of the sky (Schmitt \& Kinney 1996).  If this orientation extends down to 
kiloparsec scales, then the plane of the galactic disk would lie within the angular
range of one side of each cone (15\deg\ to 25\deg, tilted out 5\deg\ of the plane of 
the sky).  The situation then resembles NGC 1068 and NGC 4151 (Crenshaw et al. 2000,
Crenshaw \& Kraemer 2000), which also seem to have the galactic disk and one side 
of the bicone in the same plane.  Crenshaw et al. (2000) propose that the galactic
disk's ionization (by the nucleus) contributes to ENLR gas.  We propose that the
same geometry exists in this galaxy.  
 
The radio jet and the NLR emission share a similar axis, and are nearly coincident.
However, other than their spatial coincidence, there do not appear to be any other
correlations, as would be expected if the radio plasma's expansion were the source 
of the NLR velocities.  Firstly, in the data itself, there are no bright NLR clouds
that correspond to jet flux maxima.  This lack of correspondence has been noted
in other objects (NGC 4151, Kaiser et al. 2000). In addition, there is no evidence for
peculiar velocities at the positions of the radio lobes.  In terms of the dataset
velocities, there is no physical reason given by this model to explain the velocity 
trends (increasing to some turnover distance, then steadily decreasing) that we see 
in the data.  In terms of the modeling, the CT and other transverse velocity models,
predict equal blueshift/redshift amplitudes no matter what inclination angle the 
bicone is tilted.  The data show a definite difference (200-400 km s$^{-1}$) 
in velocity maxima between redshifts and blueshifts, consistent with a biconical 
geometry.  
     
\section{Conclusions}
Two STIS spectra were obtained of the NLR of the Seyfert 2 galaxy Mrk 3.  Radial 
velocities were determined of the emission-line gas as a function of position (out
to $\sim$1 kpc from the nucleus).  The velocity maps indicate general trends in the
gas motion.  These include: blueshifts and redshifts on either side of the nucleus,
steep velocity rises from systemic up to $\sim$$\pm$700 km s$^{-1}$ taking place
in the inner 0$''$.3 (0.8 kpc) both east and west of the nucleus, and gradual velocity
descents back to near-systemic values from 0$''$.3-1$''$.0.  

The data were then fitted to kinematic modeling programs for the NLR gas on the
surface of the bicone.  The data sets were fit best with a radial acceleration + 
constant deceleration model.  In the model, the cones extend out to a radius of 
0$''$.75 from the nucleus, with a half-opening angle between 15\deg\ and 25\deg.  
The modeled material reaches a maximum deprojected velocity of 1750 km s$^{-1}$, 
reaching this velocity at a distance of 0$''$.3-0$''$.43 from the nucleus, close 
to the observed distance of 0$''$.2-0$''$.3 from the nucleus.  The fit could be 
improved by positing different turnover radii and/or acceleration/deceleration laws 
for each quadrant.  Also, the high velocity data points not fit by the model appear 
to be clouds that do not encounter any dense medium and maintained their acceleration.  
Nevertheless, our goal of being able to explain all the basic trends in the data 
with a simple model was accomplished.  We have ruled out gravitational and constant 
velocity models.  We show that a model where the NLR emission is produced by expansion 
of radio jet plasma away from the radio axis does not fit the data well.

An important observational result is that the two distinct methods of obtaining 
radial velocities each gave similar results.  This has been shown previously for
NGC 4151 (Hutchings et al. 1998, Kaiser et al. 2000).  The slitless spectral procedure 
of obtaining radial velocities has proven to be a useful and efficient tool for 
quickly examining and mapping nearby galaxies with clumpy NLRs and ENLRs.  We will 
take advantage of this technique in the future to map the kinematics of the NLR in 
nine other Seyfert galaxies.

This work was supported by NASA Guaranteed Time Observer funding to the STIS 
Science Team under NASA grant NAG 5-4103 and by {\it HST}. 
Additional support for this work was provided by NASA through grant number 
HST-GO-08340.01-A from the Space Telescope Science Institute, which is operated 
by AURA, Inc., under NASA contract NAS5-26555.

\clearpage
\begin{deluxetable}{llllll}
\tablecolumns{6}
\tablecaption{Observations \label{tbl-1}}
\tablewidth{0pt}
\tablehead{\colhead{Date\tablenotemark{a}} &\colhead{Instrument} 
&  \colhead{Root Name} &\colhead{Filter/Grating} & \colhead{Exposure (s)} 
& \colhead{Slit}} 
\startdata
1991 July 18 (A)  &	PC	&W0MW0601T      &F502N   &1800	&  \\
1992 Dec 11  (A)  &	FOC	&X14W0301T      &F501N	 &1197	&  \\
1997 Oct 20  (A)  &	WFPC2	&U2E62A01T      &F606W	 &500	&  \\
2000 Jan 16  (N)  &	STIS	&O5F403010	&clear	 &20	&  open\\
2000 Jan 16  (N)  &	STIS	&O5F403020	&G430M   &2154	&  open\\
2000 Aug 22  (N)  &	STIS	&O5KS01010	&G430L	 &1080	& 52$''$x0$''$.1 \\
\enddata
\tablenotetext{a}{A-Archival, N-New}
\end{deluxetable}

\clearpage
\begin{deluxetable}{ll}
\tablecolumns{2}
\footnotesize
\tablecaption{Modeling Parameters \label{tbl-2}}
\tablewidth{0pt}
\tablehead{\colhead{Parameter}  &   \colhead{Symbol or constant value}}
\startdata
Min \& Max distance of cones (pcs)  &   Min. = 0 , Max = D    \\
Min \& Max half-opening angle  &   $\theta$$_{min}$, $\theta$$_{max}$  \\
Inclination angle	       &   $i$ \\
Deprojected Maximum velocity 
of NLR gas (km s$^{-1}$)   &   V$_{max}$	 \\
Velocity Laws		       &   \it Constant velocity (CV)   \\  
			       &   \it Radial Acceleration (RA)   \\
		 &   \it Radial Acceleration plus Constant Deceleration (RA+CD)  \\
			       &   \it Constant Tangential Flow (CT)   \\
			       &   \it Gravitational Infall (GI)   \\
Center of Slit 		       &   Centered on optical continuum peak  \\
Position angle of the long slit	 &    $\sim$71\deg    \\
Slit Width (in pixels)	       &   0$''$.1   \\
\enddata
\end{deluxetable}

\clearpage
\begin{deluxetable}{lllll}
\tablecolumns{5}
\footnotesize
\tablecaption{Parameters of each best fit model}
\tablewidth{0pt}
\tablehead{\colhead{Parameter} &  \colhead{CV} &  \colhead{RA}  &  \colhead{RA+CD} & 
\colhead{CT}}
\startdata
D (pcs)	&   80	&   80    &     80	&   80	\\
$\theta$$_{min}$, $\theta$$_{max}$  &  15, 25  &   15, 25   &   15, 25	 &  15, 25\\
$i$	&   5\deg  &  5\deg   &  5\deg   &   5\deg	\\
V$_{max}$ (km s$^{-1}$) &  1400	&   3000   &   1750   &  550  \\
\enddata
\end{deluxetable}

\clearpage
\figcaption[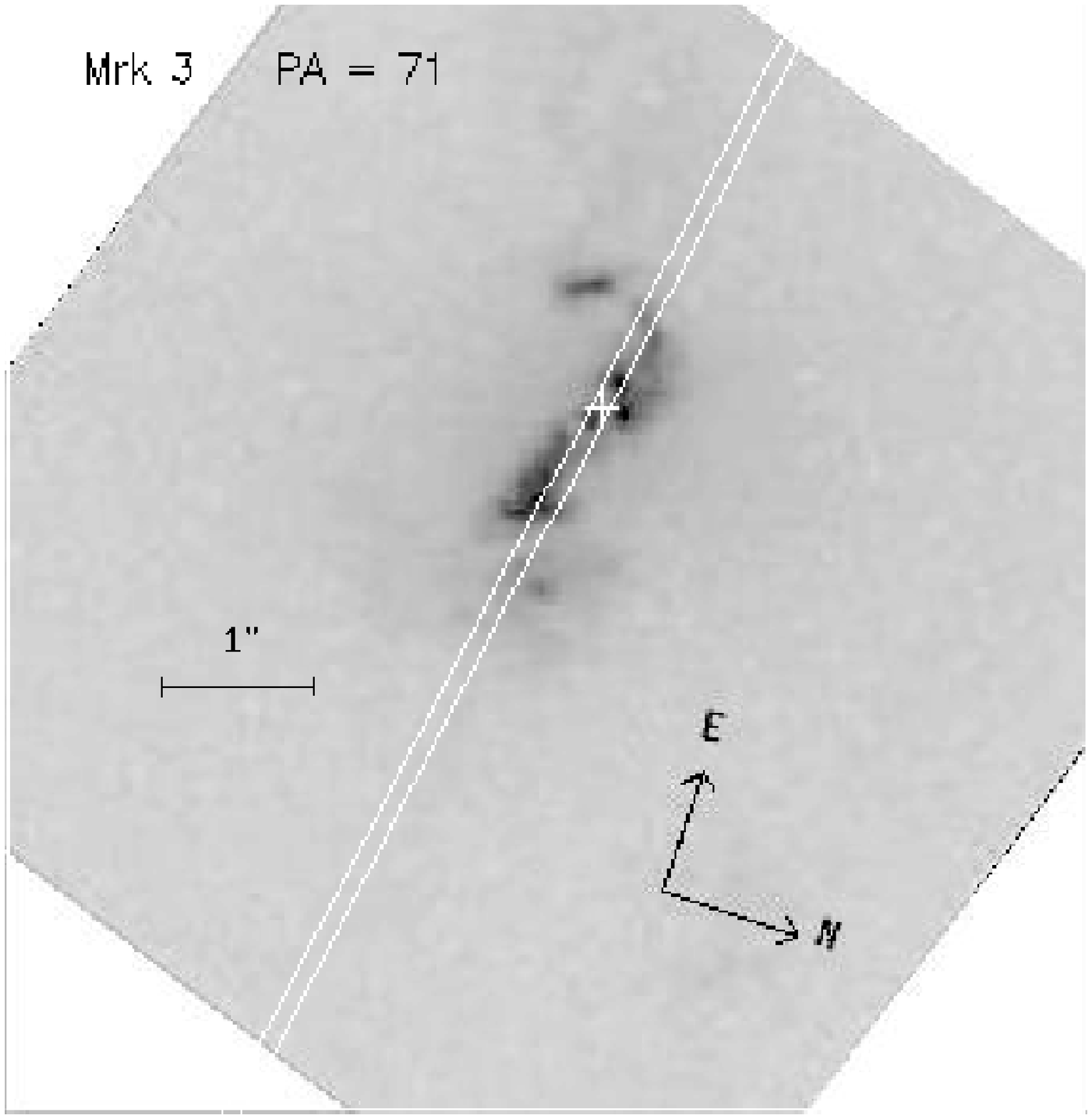]{FOC image of the bright NLR clouds of Mrk 3.  The 
position of the long slit is seen to pass through the nucleus, and through
the clouds in spots.  The cross depicts the position of the nuclear continuum
center.  The backward `S' shape of the main clouds is seen.}

\figcaption[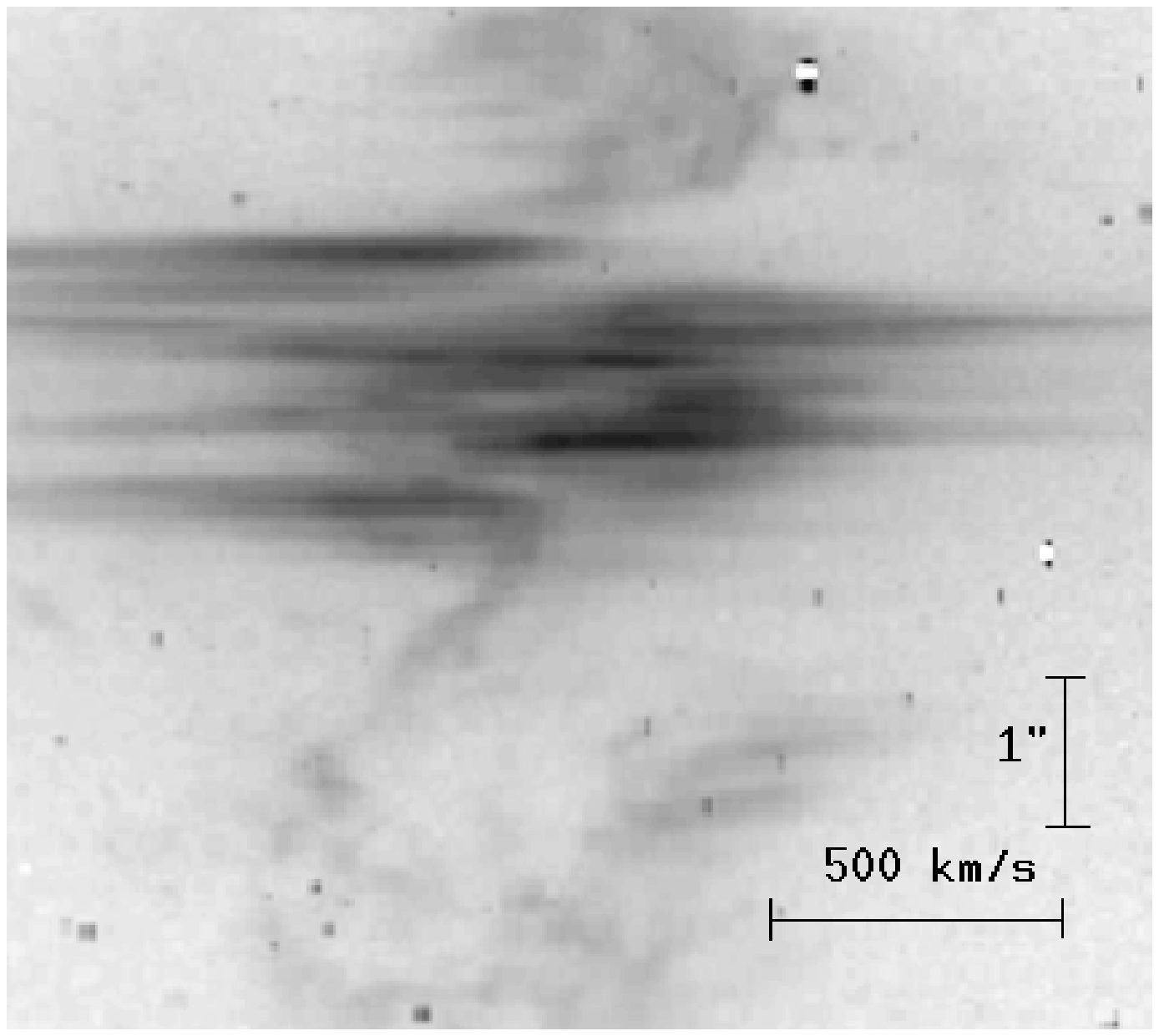]{STIS slitless spectrum showing the region around 
[O III] $\lambda$5007.  The horizontal axis is along the dispersion, while the
vertical scale is the spatial axis.  Note the high dispersions of the NLR clouds.
The clouds are also shifted slightly along the dispersion axis, the shift
allowing the radial velocity to be calculated.  The fainter ENLR clouds can
also be seen $\sim$1'' above and below the NLR.  Their dispersion can also be
seen.}

\figcaption[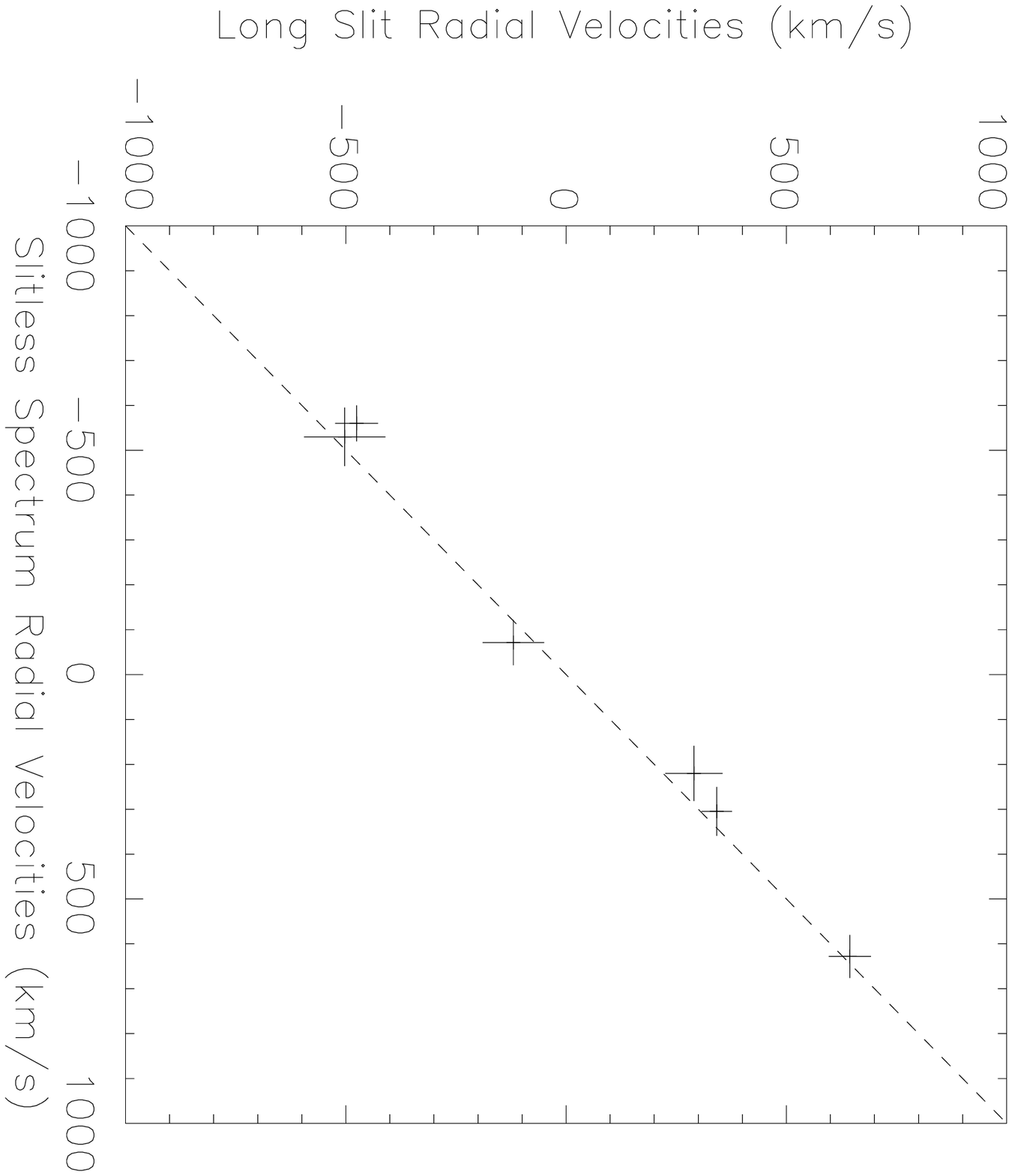]{Radial velocities averaged over bright clouds in the long slit 
are shown plotted against their corresponding values obtained with the 
slitless spectrum.  The error bars represent the standard deviations from 
the averages.} 

\figcaption[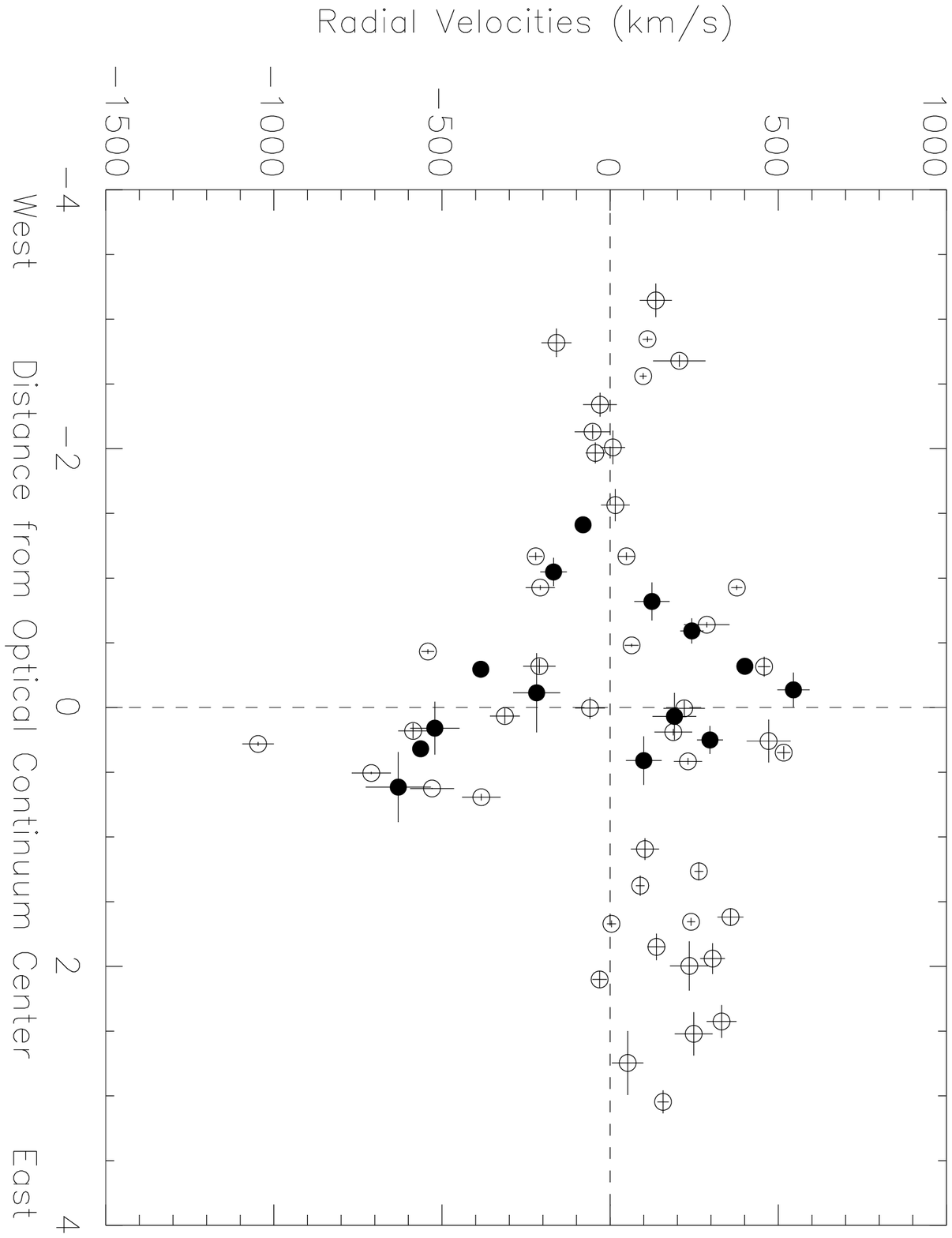]{Radial velocities (with respect to systemic) 
averaged over each individual cloud vs each cloud's distance from the optical 
continuum center.  Standard deviations in the position and velocity averages 
are shown as the error bars.  The slitless clouds (marked with open circles) 
include all clouds, including those not lying in the slit, and those extending 
out into the ENLR (from 1$''$.5 west of the nucleus outwards, and from 1$''$.0 
east of the nucleus outwards).}

\figcaption[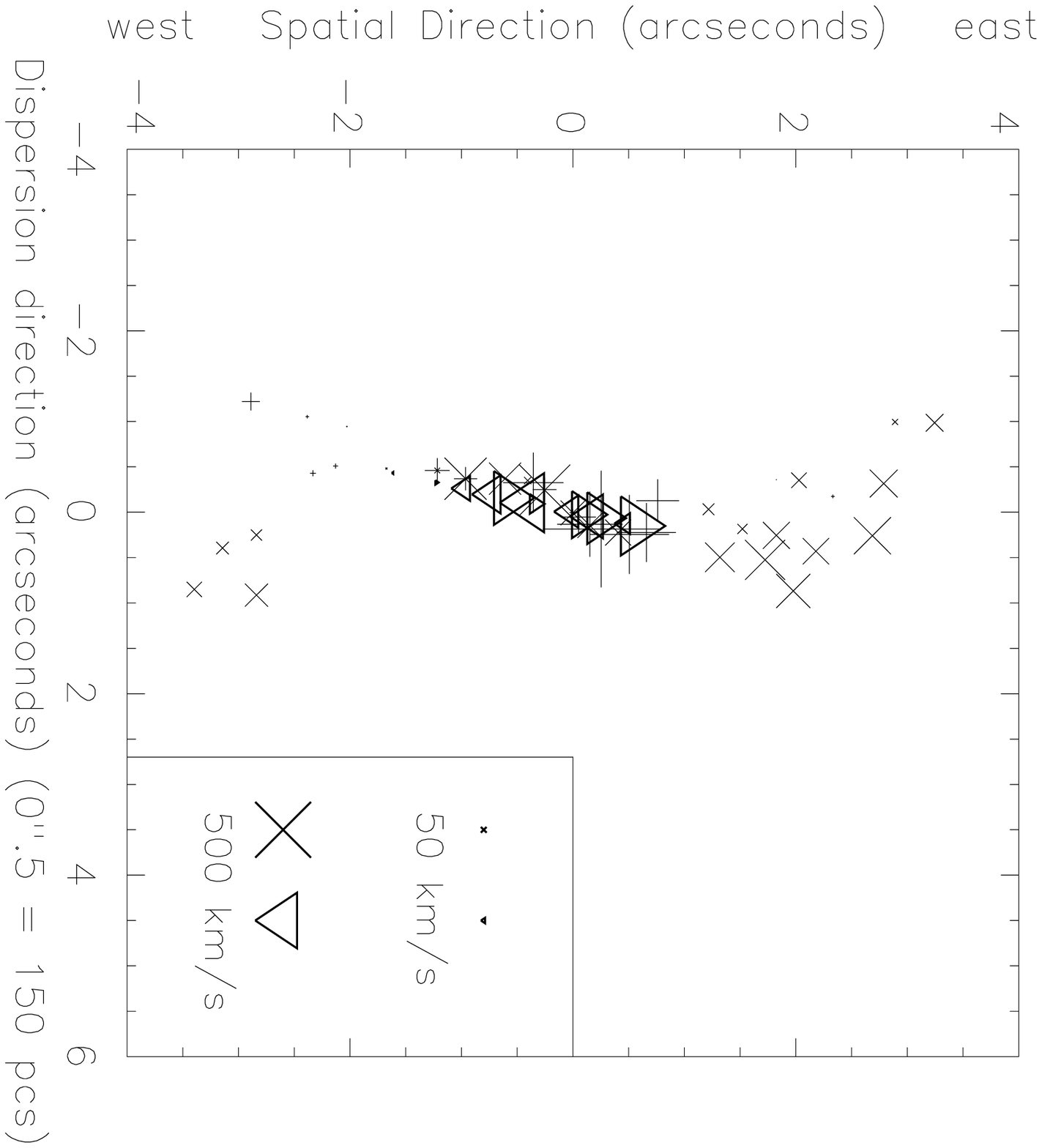]{This figure shows the positions in the sky of the 
clouds, with their radial velocity magnitudes symbolised by the size of the 
figure used to identify them.  The blueshifted clouds are symbolized by a 
``+'', while redshifted clouds are an ``x''.  The radial velocity measurements 
are binned over each cloud.  The ENLR clouds can be seen to be mostly 
redshifted.  The long slit clouds are differentiated by being in boldface.}

\figcaption[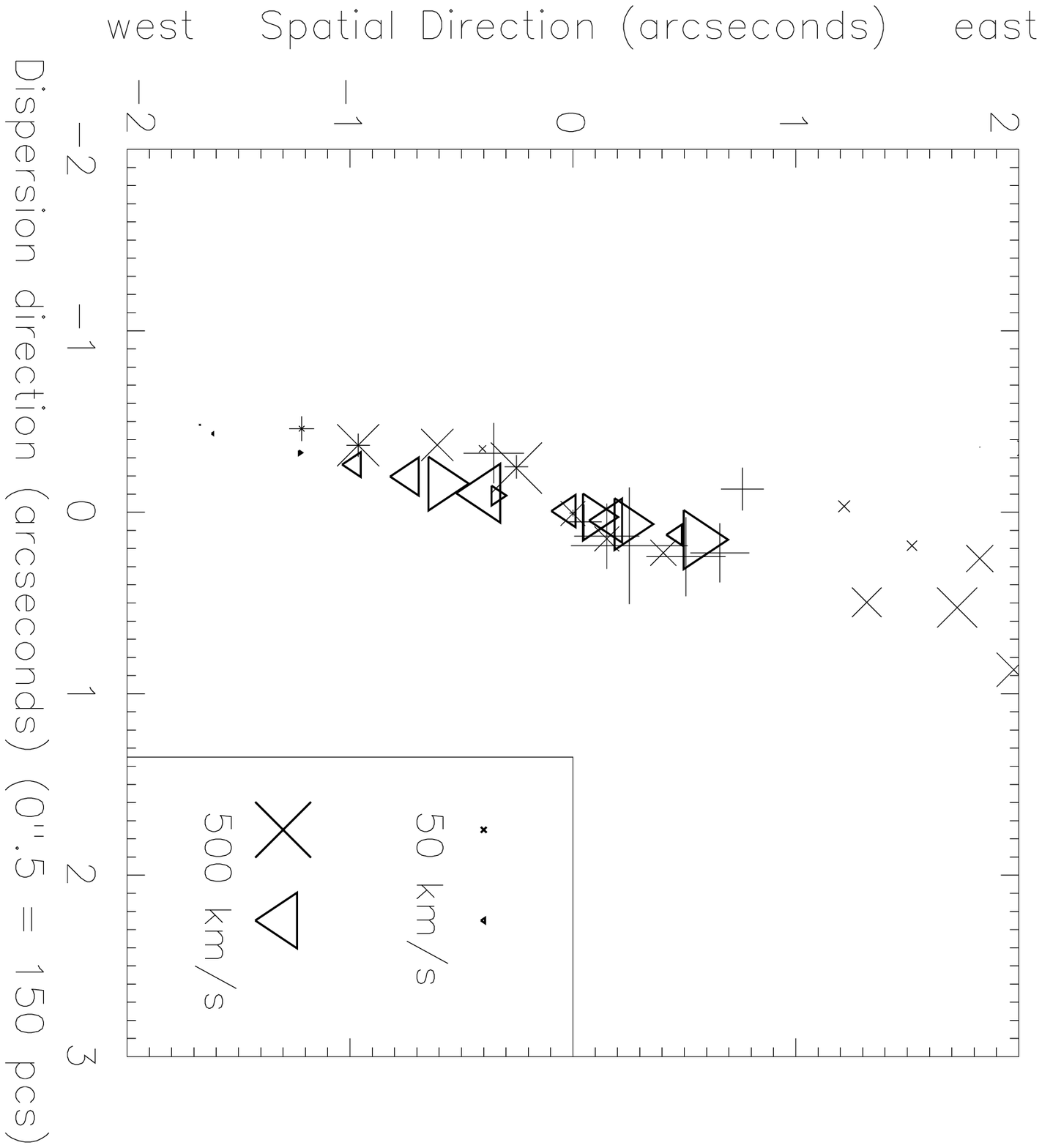]{Same as Figure 5, but zoomed in on the 
NLR clouds.}

\figcaption[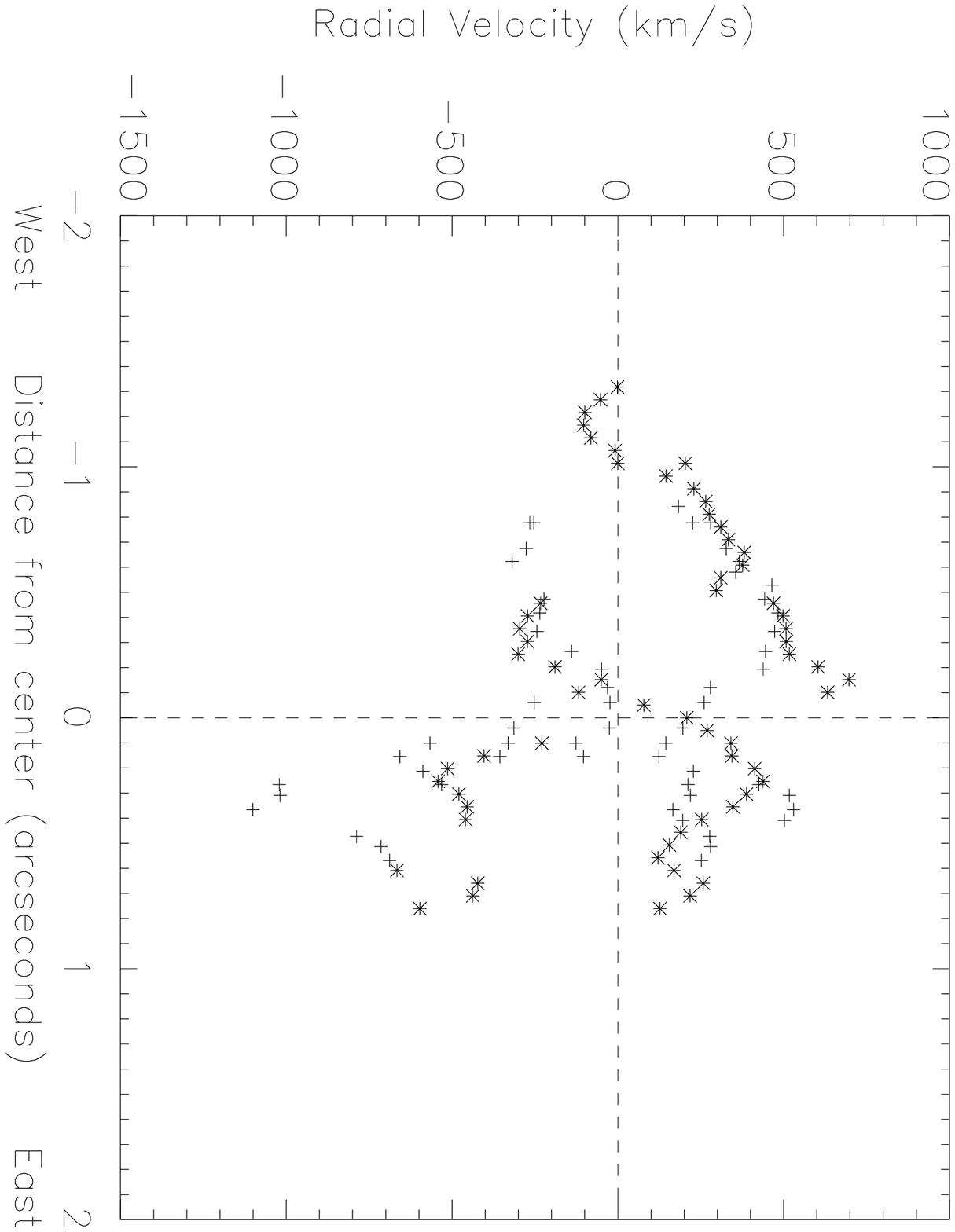]{Unbinned radial velocity of each row 
perpendicular to the dispersion direction plotted against the distance 
from the center.  The slitless spectral data points plotted here are only 
those observed to lie within the slit.  Errors are roughly equal to the 
size of the symbols.}

\figcaption[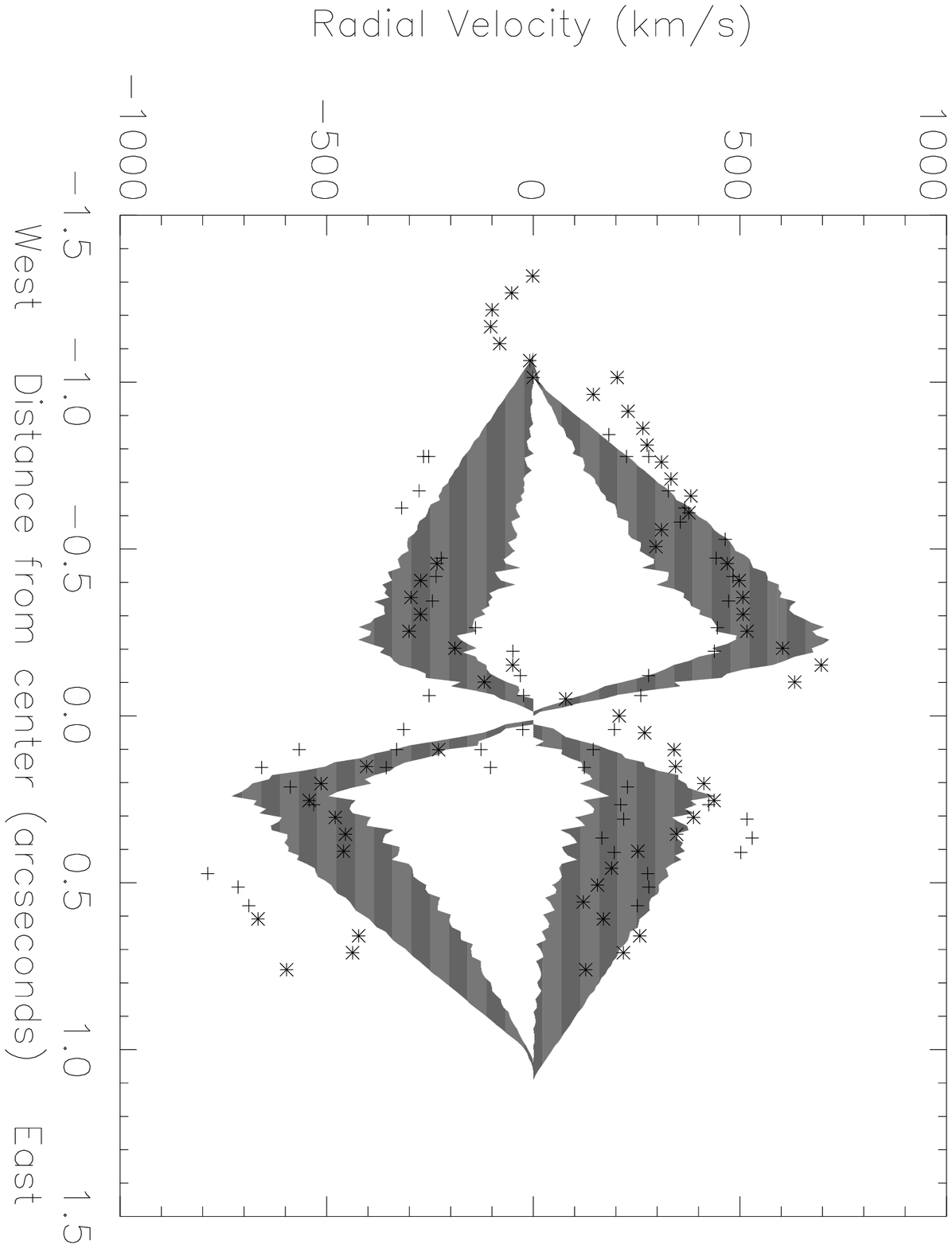]{Same as Fig. 7, but overlaid with the best fit model
of the radial acc + constant decel model.  Parameters of the model are given
in Table 3.}

\clearpage
\begin{figure}
\plotone{ruiz.fig1.ps}
\\Fig.~1.
\end{figure}

\clearpage
\begin{figure}
\plotone{ruiz.fig2.ps}
\\Fig.~2.
\end{figure}

\clearpage
\begin{figure}
\plotone{ruiz.fig3.ps}
\\Fig.~3.
\end{figure}

\clearpage
\begin{figure}
\plotone{ruiz.fig4.ps}
\\Fig.~4.
\end{figure}

\clearpage
\begin{figure}
\plotone{ruiz.fig5.ps}
\\Fig.~5.
\end{figure}

\clearpage
\begin{figure}
\plotone{ruiz.fig6.ps}
\\Fig.~6.
\end{figure}

\clearpage
\begin{figure}
\plotone{ruiz.fig7.ps}
\\Fig.~7.
\end{figure}

\clearpage
\begin{figure}
\plotone{ruiz.fig8.ps}
\\Fig.~8.
\end{figure}

\pagestyle{empty}
\end{document}